# Real-Time Parallel Programming:
# State of Play and Open Issues

Luis Miguel Pinho

*School of Engineering of the Polytechnic of Porto (ISEP) & INESC TEC*

*Porto, Portugal*

**Abstract**

*Real-time systems applications usually consist of a set of concurrent activities with timing-related properties. Developing these applications requires programming paradigms that can effectively handle the specification of concurrent activities and timing constraints, as well as controlling their execution on a particular platform. The prevailing trend for high-performance, and the use of fine-grained parallel execution makes this an even more challenging task.*

*This paper provides an overview of the state-of-the-art and challenges for the development of real-time parallel applications, focusing on two current research directions, one from the high-performance arena (based on OpenMP) and another from the real-time and critical systems domain (based on Ada). The paper provides a review of the supported features of each one of the approaches, identifying the still open issues and the current research directions.*

## 1. Introduction

Real-time systems are those systems where, apart the functional requirements, the computing system needs to provide results within specific time intervals (deadlines), emanating from application requirements [1]. These systems are an important and challenging area, with a wide spectrum of applications, from industrial automation to the Internet-of-Things, being a cornerstone of "cyber-physical systems": in most of these applications, the computing system interacts, sensing and actuating, with the external environment [2]. In this case, the timing correctness of the results of the computing system can be critical to the correct functioning of the physical environment.

The criticality of the deadlines is basically categorized into hard real-time and soft real-time [1]. In a hard real-time deadline, not providing the result within the required interval, implies that the computing system failed, whilst for soft real-time, a missed deadline can lead to degraded, but still acceptable, behaviour. Most cyber-physical systems have hard real-time requirements, in which some deadlines must be strictly met. And in most cases (e.g., avionics, automotive), cyber-physical systems are used in critical applications, therefore failure of a deadline may have severe consequences.

---

[1] Other categories exist, but for the majority of the cases, differentiating between hard and soft is sufficient.

A typical characteristic of real-time systems is the interplay of events and physical-related activities [2] that occur sequentially or concurrently under these timing constraints. Activities can be periodic or non-periodic (i.e., event-driven). Periodic activities repeat regularly after a fixed time interval. Non-periodic activities are further divided into sporadic, where a maximum frequency of repetition can be derived, and aperiodic, where the release frequency is not bounded [3].

To allow analysing if the application deadlines can be met, one of the challenges is to determine the response time (the time interval between a stimulus that triggers a computing activity and the output of the result or action) for the application activities, in the particular platform to be used in the system. The goal is to obtain a worst-case value, which is a safe upper bound for all executions of the application. Although this could be eventually simple, the interference between the concurrent activities, and the increasing complexity of hardware architectures provides significant challenges.

The methods used to execute these activities in the computing system are therefore of paramount importance [3], with numerous real-time computing mapping and scheduling approaches being proposed to guarantee that results are provided in a bounded and known time (in principle, before deadlines). As a complement to the mapping and scheduling of computation, the real-time scheduling theory focuses on designing application, scheduler, and platform models and developing tools and techniques that together, allow for the time behaviour of the entire system to be anticipated, modelled and analysed. These tools and techniques can be broken in two broad and distinct categories: timing analyses and schedulability analyses.

In timing analysis, the objective is to compute tight bounds (or probabilistic profiles) on the time needed to perform an operation executed in isolation, e.g., to determine the worst-case execution time (WCET) of the activity [4]. In schedulability analysis, the objective is to check analytically, at design or run time, whether all the timing requirements of the system will be met [3], considering the WCET and the interplay and interference between activities.

The methodology for the design of real-time systems has been challenged due to the increasing processing requirements of modern cyber-physical systems [5]. The current trend to increase processing power by manufacturing chips including multiple processor cores has popularised the ability to execute the inherent concurrent activities using fine-grained parallelism. This cheaply available computational power already makes parallel programming a concern for software developers, since the sequential programming model does not scale well for such multi-core systems [6]. Adding to this the constraints introduced by the timing requirements and the close interaction with the physical world makes this problem even more difficult [7].

One of the areas where the addition of fine-grained parallelism has also impacted the development of real-time systems, is the required abstractions and mechanisms provided to programmers for the development of software applications. Real-time programming paradigms are required to provide capabilities to express and control the concurrent and parallel execution of the activities, in order to enable the development of safe (from a timing perspective) system.

---

[2] In the real-time systems domain, the term "task" is often used to denote such computing activities. However, the term "task" is significantly overloaded in computing, with relevant other uses being:
  - In OpenMP, tasks denote the specific instances of executable code and data environment, which are to be executed by threads.
  - In Ada, tasks are the language entities for concurrent programming (which can be used to implement real-time tasks).
  - In operating systems, task may represent thekernel structure that specifies a thread.

This paper provides the current state-of-the-art in the integration of parallel programming within real-time systems. The focus is the classic real-time programming paradigm [8], also denoted as the asynchronous paradigm [9] or the scheduled model [10].

The paper is structured as follows. The next section provides an overview of the topics of real-time systems and parallel models, and how parallelism impacts in the development of real-time systems. Afterwards, Section 3 provides a brief description of real-time programming approaches, using examples from both the POSIX specification and the Ada programming language.

Section 4 then sets forth the requirements on models for programming parallel real-time systems, with two different approaches being developed to support programming parallel real-time discussed in Sections 5 and 6. Section 5 focuses on the use of the OpenMP specification, in particular the OpenMP Tasking Model, while Section 6 focuses on the forthcoming revision of the Ada language standard. The two examples provide different perspectives to the problem of integrating parallelism and real-time: OpenMP fully supports parallel programming, with current activities being on it can be also used for real-time systems (potentially with POSIX support); Ada, on the other hand, already provides extensive support for programming real-time systems, with parallelism being now provided in the new version being standardised (potentially on top of OpenMP runtime).

## 2. Real-Time and Parallel Models

The basic real-time systems computing model is of an application constituted by a set of recurrent activities executing in the same hardware platform (single or multicore), competing for the same resources, using a well-defined approach to manage their execution on the platform (a single processor or a multicore). Scheduling algorithms range [3] from static allocation of priorities to activities (fixed priority scheduling) to computing during execution the activity with highest priority in the processor (being the most known example earliest deadline first scheduling). Also, existent approaches consider both pre-emptive and non-pre-emptive scheduling of activities (in the former, if an activity with higher priority becomes ready to execute, it pre-empts the activity being executed) as well as more cooperative scheduling, such as server-based or round-robin (activities time share the processor with pre-defined budgets). Figure 1 provides an example with 3 activities executing in a single processor, using a fixed priority pre-emptive scheduling [3].

---

[3] The vertical ordering of activities is many times used to denote the priority ordering of the activities (top meaning higher priority).

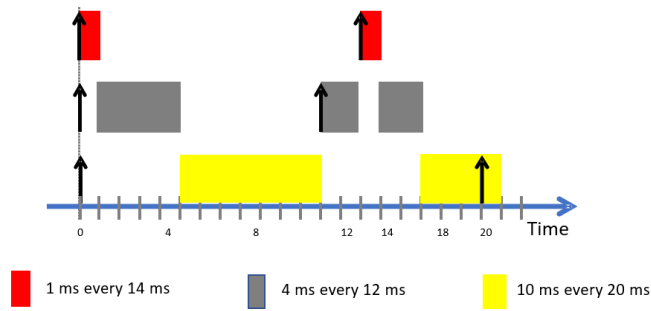

**Figure 1.** Fixed priority scheduling.

The fine-grained parallel programming model allows programmers to concentrate on exposing the available parallelism and express it within the concurrent activities, with the parallel execution being dynamically managed by a run-time system. This implies that one activity no longer specifies a sequence of code, but instead a graph of computation, with a set of communicating code segments, which may fork or join communication paths.

Within the real-time systems domain, and in order to characterize more in detail the parallel activity structure, the following models have been proposed (Figure 2):

- Fork/join model [11], where each activity is divided into sequential and parallel segments. Parallel segments must be preceded and followed by a sequential segment, and the number of threads cannot be greater than the number of processors in the platform.

- Synchronous parallel model [12], which generalizes the fork/join model by considering activities composed of different segments that may have an arbitrary number of parallel threads (even greater than the number of cores).

- Parallel DAG model [13], further generalizing the model by identifying each activity with a directed acyclic graph (DAG), where each node represents a sequential code segment, and each directed edge represents a precedence constraint between two segments.

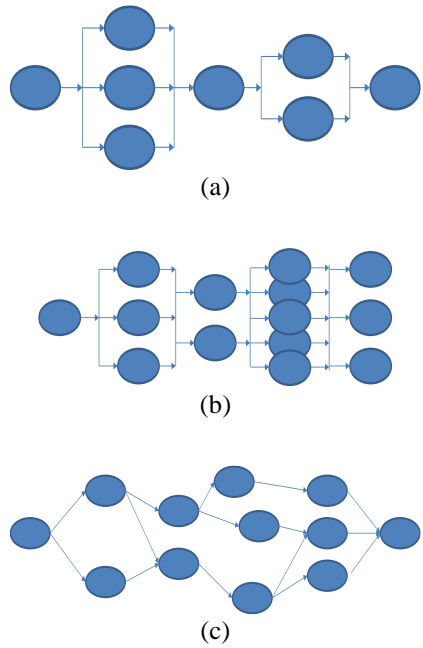

**Figure 2**. Fork/join (a), Synchronous (b) and parallel DAG (c) models.

Note that the execution time of an activity is now a set of segments' execution times, and the response time is now defined as the end-to-end response time of the graph. Each model provides different methods to evaluate if the end-to-end response time allows, in the worst-case, the execution to fulfil the timing requirements (the deadline) of applications.

## 2.1 Mapping and Scheduling Parallel Computation

In order to be able to meet the timing requirements of parallel applications, it is necessary to provide the system with appropriate means to map the computing activities to the underlying operating system threads (mapping), and dynamically schedule these threads to achieve both predictability and high-performance (scheduling).

The timing behaviour of an application is significantly affected by the mapping of the parallel code segments to cores in a parallel platform, due to the interference of several activities mapped in the same cores, the interference between different cores accessing shared platform resources (e.g., memory) and the sharing of data between activities.

Figure 3 (a) shows a simple example with two activities executing in a platform with two processors (using fixed-priority pre-emptive scheduling). A simple mapping would be to map activity 1 to processor 1, except the parallel segment 'c', which is mapped to processor 2 (to allow for parallelism), and activity 2 to processor 2, except for the parallel segment 'g'.

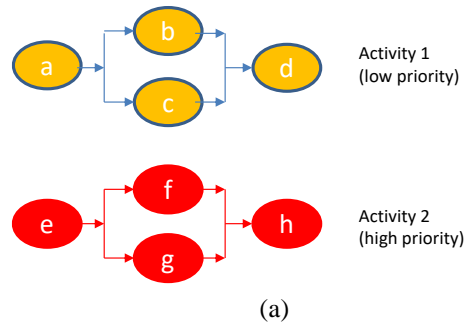

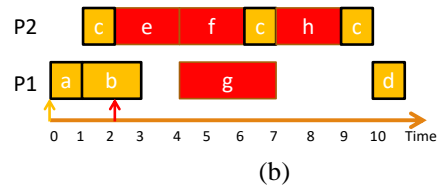

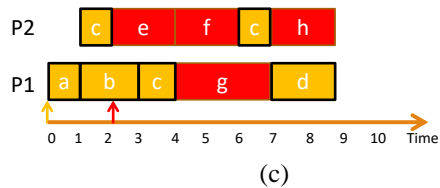

**Figure 3**. Mapping example: (a) activities; (b) mapping with no migration; (c) mapping with migration.

However, as can be seen in Figure 3 (b), if activity 2 is released at instant 2, there will be two slots where there is a processor (P2) that cannot execute although there are pending activities. At instant 3, P2 cannot execute segment 'd', as 'c' is not finished, and cannot execute 'g', as 'e' is also not finished [4]. At instant 7, 'h' pre-empts 'c' in P1, and P2 still cannot execute 'd'. If 'c' is allowed to migrate between processors (Figure 3 (c)), the empty slots can be removed, as P2 can execute 'c' at instants 3 and 7, reducing the overall response time.

As seen in Figure 3, the mapping can be dynamic, i.e., decided during execution (c), or static, i.e., "forced" before execution (b). Dynamic mapping can lead to better average performance, whilst static mapping may provide better predictability (jitter between best and worst case is reduced) [14]. Dynamic mapping, if also supporting migration, as in Figure 3 (c), can also incur an overhead, as the migration of execution between processors is not without computation cost.

The scheduling of a set of non-parallel activities in parallel platforms has been the subject of extensive research in the real-time systems community [15]. For hard real-time applications on multiprocessors, partitioned scheduling prevails, wherein subsets of activities are assigned to specific processors. This allows the use of well-performing, practical-to-implement and long-standing uniprocessor scheduling algorithms. It also largely provides temporal and functional isolation for activities on different processors, except for interactions with shared resources.

---

[4] This problem also occurs for non-parallel activities, when multiple processors are considered. However, it is deeply exacerbated by parallelism, which also explodes the number of different mapping and scheduling possibilities that can be considered.

However, as seen in Figure 3 (b), partitioning often wastes processing capacity via fragmentation. Conversely, fully-migrative scheduling, which allows activities to migrate between processors even halfway through their execution, overcomes such utilisation issues but brings implementation and synchronisation overheads, break of temporal isolation and difficulty in analysis. The interest in hybrid "semi-partitioned" schemes, under which only a subset of the activities migrate – and only in a very controlled manner, aims for the best of partitioned and migrative scheduling.

For real-time applications structured using fork/join or synchronous parallel models, the parallel code segments can be mapped to a set of operating system threads, which are either mapped on the various processors as independent entities with no precedence constraints or globally scheduled, with the subsequent segment only allowed to continue (in a single thread) after all such parallel segments complete [11]. Extensions [12][13] allow for directly alternating different parallel phases (with different numbers of threads), without intermediate sequential phases. In this case, different mappings of the parallel tasks to the threads can be done, even with static mapping approaches, in order to increase system utilisation whilst maintaining predictability [16][17]. More dynamic approaches can also be used [18], where real-time activities also employ work-stealing [19] for load balancing, using dynamic mapping in this case, with the real-time scheduler allowing code blocks to migrate between threads when cores are idle.

## 2.2 Timing and Schedulability Analysis of Parallel Real-Time Applications

In parallel processors, the problem of finding WCET upper bounds is complicated by the high number of resources shared between the cores [20]. In such platforms activities may execute concurrently on different processor cores but any of their accesses to a memory may contend with others, emitted by activities running on other cores. On complex architectures the time during which a core stalls waiting for a memory or bus request to be served is a significant component of the overall execution time of a program, and may vary considerably, particularly when code blocks may be dynamically mapped into the different cores [21].

The area of schedulability analysis of parallel models is currently a major topic in the real-time systems community (some examples are [11][12][13][22][23]), but despite the vast amount of research conducted to explore techniques that allow achieving a predictable behaviour on top of parallel platforms, the state-of-the-art in timing and schedulability analysis is still far from ready solutions. Problems lie in the characterisation of the multiple interferences and inter-dependencies that may arise among the various hardware components in a parallel system. When multiple activities execute simultaneously on the same processor, or even in separate processors they may experience significant contention accessing shared resources, like shared memories, communication networks or acceleration devices. Such contention has to be considered and factored in the timing and schedulability analysis, potentially leading to very pessimistic response-time bounds. Indeed, worst-case scenarios that take into account corner cases, e.g., all cores accessing the same resource at the same time, may cause a significant inflation of the execution and response times, or be intractable in practice [14]. This inflation increases exponentially with the number of cores, requiring a great share of processing to be guaranteed for these rare cases, largely under-utilising the system. Therefore, and in order to reduce such pessimism, practical approaches tend to perform static mapping of parallel computation to partitioned scheduling systems, improving overall guaranteed utilisation [16][17].

The challenges of mapping and scheduling parallel computation, and the associated timing and schedulability analyses, is a very rich research topic, which is nevertheless orthogonal to the subject of this paper. Instead, the paper focuses on the also important aspect of the capabilities which are provided to the programmer, to be able to express and control the concurrent and parallel execution of the activities, in order to enable the development of safe (from a timing perspective) system.

## 3. Programming Real-Time Systems

In order to address the specific requirements of real-time systems, programming models and paradigms usually provide a set of abstractions which intend to capture the real-time systems scheduling model [8][9][24]:

- Notion of time:

  In real-time systems, representation and accuracy of the time base is fundamental.

- Specification of real-time activities:

  As much as possible the ability to specify recurrent time-drive (periodic) or event-driven (sporadic or aperiodic) activities in the programming model.

- Specification of activity properties:

  In the real-time system domain, to guarantee that activities meet the deadlines of controlled systems, scheduling approaches are based on properties such as priority, deadline, execution time, therefore the programming model should support these properties.

- Communication and synchronization

  Providing mechanisms that allow activities to share data and synchronize between concurrent activities, with timing constraints.

- Mapping of computation:

  The ability to control how activities are mapped to the underlying processors, either by pinning to cores, or allowing activities to migrate within, or across, processors.

- Control of activity scheduling:

  Guaranteeing that the execution of the concurrent activities meets the model considered for schedulability analysis requires the ability to control how concurrency is scheduled in the underlying platform, e.g., using pre-emptive or non-pre-emptive scheduling, or fixed priority, earliest deadline first or round robin scheduling approaches.

To support these requirements there are basically two different approaches:

- Use a sequential programming language with the concurrency and real-time requirements being met by libraries and operating system calls: this is one of the most common approaches, using the

C [25] language and real-time operating systems, eventually with the POSIX real-time support [26].
- Use of a language with support for concurrency and real-time as language first-class entities, such as Ada [27] or Real-Time Java [28].

These two approaches are briefly described in the remainder of this section with examples from C/POSIX and Ada. The goal is to provide a brief overview of the approaches, and not provide a full coverage of the support. Readers are referred to [24] and the specifications [26][27] for more detailed information.

## 3.1 Real-Time Support in the POSIX Specification

The POSIX specification provides a significant number of mechanisms which support programming real-time systems. The specification of the concurrent activities can be done using pthreads, as shown in Listing 1, in this case an example of specifying a periodic activity.

```c
void *activity_in_thread_code(void *arg) {
   long period_us *ps = (long *) arg;
   struct timespec next, now;
   clock_gettime(CLOCK_REALTIME, &next);
   while (1) {
      // activity code

      timespec_add_us(&next, *ps);
      clock_nanosleep(CLOCK_REALTIME,TIMER_ABSTIME,&next,NULL);
   }
   return NULL;
}
```

**Listing 1**. Example of programming a periodic activity.

Note also the use of the `clock_nanosleep` function, together with a real-time clock (`CLOCK_REALTIME`) and the specification of the absolute time flag (`TIMER_ABSTIME`). Supporting the notion of time is fore sure of paramount importance for real-time systems. More than considering wall/calendar type of clocks, real-time systems require a time base which is precise and progresses at a constant rate and is not subject to the insertion of extra ticks to reflect leap seconds (as calendar clocks are). A constant rate is needed for control algorithms which required to be executed on a regular basis. Strangely, this is not the case of the `CLOCK_REALTIME` time base, as it may change with changes to the operating system wall time, thus POSIX also provides a `CLOCK_MONOTONIC` time base which is intended for this purpose [5] [26] (section B.2.8).

In order to implement a periodic activity, sleep operations should use absolute time, and not relative time. The reason is that the thread can be interrupted immediately before the call to the sleep function. If sleep was relative, e.g., sleep for 5 seconds, the next release of the periodic activity would be variable and depending on the eventual interruption.

---

[5] Note that the Linux implementation of CLOCK_MONOTONIC does not fully adhere to POSIX, and in Linux CLOCK_BOOTTIME should be used instead.

This handling of time may be complemented (optional feature) with the ability to use CPU execution time clocks, which measure the amount of CPU time that is being consumed by a particular thread, and setup execution time timers, which are able to measure (and notify) of threads overrunning their "worst-case execution time" [6]. This is based on the existence of per process and per thread CPU-time clocks (`CLOCK_THREAD_CPUTIME_ID`), and the possibly to create a timer that calls a handler function (using signals) when the time expires (Listing 2). The thread can eventually poll the execution time with `timer_gettime`, as an alternative method to detect overruns.

```
void sig_handler(){
    // process execution time overrun
}
//...
struct sigevent event;
timer_t timer_id;
struct itimerspec cpu_usage;

// handler function
SIGEV_SIGNAL_INIT(&event, SIGALRM);
signal(SIGALRM, sig_handler);

// Create a thread timer
timer_create(CLOCK_THREAD_CPUTIME_ID, &event, &timer_id);

// Set execution time
cpu_usage.it_value.tv_sec = 0;
cpu_usage.it_value.tv_nsec = exec_time;
cpu_usage.it_interval.tv_sec = 0;
cpu_usage.it_interval.tv_nsec = period;

// Set the timer
timer_settime(timer_id, 0, &cpu_usage, NULL);
```

**Listing 2**. Setting an execution time timer.

The POSIX specification also provides a limited approach to specify real-time properties and scheduling, basically supporting fixed priority and round robin scheduling, and the definition of thread priorities (Listing 3). An optional policy exists, which implements the notion of a sporadic server [29], an approach that provides pre-allocated periodic server budgets to account for aperiodic activities. Nevertheless, this policy is not implemented in the large majority of the systems.

```
pthread_attr_t custom_attr_fifo;
int fifo_max_prio;
struct sched_param fifo_param;

pthread_attr_init(&custom_attr_fifo);
pthread_attr_setschedpolicy(&custom_attr_fifo, SCHED_FIFO);
fifo_param.sched_priority = sched_get_priority_max(SCHED_FIFO);
pthread_attr_setschedparam(&custom_attr_fifo, &fifo_param);
pthread_create(&thread_id, &custom_attr_fifo, thread_func,NULL);
```

**Listing 3**. Example of managing scheduling properties with POSIX.

Note that a system is not restricted to only implement the mechanisms of POSIX. For instance, the Linux kernel also implements the `SCHED_DEADLINE` policy (Listing 4), which supports a more dynamic real-time scheduling algorithm, Earliest Deadline First, also with budgets (using a variant of the Constant Bandwidth Server [30]). And the specification defines the possibility of a `SCHED_OTHER` class, where an operating system can implement its own scheduling, at the expense of the code losing portability across operating systems.

---

[6] Overrunning worst-case execution time may occur both due to erroneous situations (e.g., a livelock), or due to incorrect calculation (or estimation) of the execution time of the code.

```
struct sched_attr attr;
attr.sched_policy = SCHED_DEADLINE;
attr.sched_runtime = 10 * 1000000; // 10 ms
attr.sched_period = attr.sched_deadline = 100 * 1000000; // 100 ms
sched_setattr(0, &attr, flags);
```

**Listing 4**. Specifying a deadline with budget for the current pthread in Linux.

POSIX also supports mechanisms (Listing 5) to address potential priority inversion [31] when synchronising threads using mutexes. Priority inversion occurs when a high priority thread is waiting for a mutex, which is locked by a low priority thread, and the low priority thread is not able to execute due to threads with intermediate priorities. Common approaches to address this problem rely on priority inheritance (the low priority thread inherits the priority of the blocked high priority thread) or priority ceiling (the low priority thread executes at the highest priority of any thread that can request the mutex). Although POSIX specifies both priority inheritance and priority ceiling protocols, most of the implementations only support the first.

```
pthread_mutexattr_t attr;
pthread_mutex_t m;
pthread_mutexattr_setprotocol(&attr, PTHREAD_PRIO_INHERIT);
pthread_mutex_init (&m, &attr);
```

**Listing 5**. Specifying a mutex with priority inheritance.

Concerning the ability to map computation to the underlying processors, although not part of the POSIX specification, many implementations of the pthread library provide mechanisms to define the set of cores (CPUs) where a thread may execute, using affinities. Listing 6 provides an example where the thread executing the code will be only allowed to execute in cores 0 and 1 of the processor.

```
cpu_set_t cpuset;
pthread_t thread;
//...
thread = pthread_self();
CPU_ZERO(&cpuset);
for (int i = 0; i < 2; i++)
    CPU_SET(i, &cpuset);
pthread_setaffinity_np(thread, sizeof(cpuset), &cpuset);
```

**Listing 6**. Linux non-portable setting of affinities.

The use of a sequential language combined with library and real-time support introduces several issues [24]. The first one is that the compiler (and even the operating system) is not aware of the concurrency between the different threads, therefore most of the concurrency errors cannot be easily detected, particularly in the compile phase. And the readability of the code is highly impaired, since the concurrent nature of the application, and eventual communication/synchronization, is not directly visible. Also, as seen in the previous paragraphs, even with existing specifications, programs are many times not portable across different operating systems.

## 3.2 Real-Time Support in the Ada Language

An alternative approach is to use a language which supports as much as possible concurrency and real-time concepts as first-class language entities. One of such examples is the Ada language [27]. The Ada standard specifies a set of core mechanisms, of which Chapter 9 relates to concurrency, as well as a set of Annexes which may be optionally provided. In particular, Annex D specifies a set of mechanisms to support real-time programming.

The Ada language provides language support to concurrent activities by a language type `task`. Task objects are logical autonomous concurrent activities, with independent state. The language does not force how the runtime maps the tasks to underlying threads, as long as the concurrency properties are guaranteed [32] (Chapter 9, paragraph 11):

"*Concurrent task execution may be implemented on multicomputers, multiprocessors, or with interleaved execution on a single physical processor. On the other hand, whenever an implementation can determine that the required semantic effects can be achieved when parts of the execution of a given task are performed by different physical processors acting in parallel, it may choose to perform them in this way.*"

The language provides a model for the specification of periodic and aperiodic/sporadic tasks, although user programmed (Listing 7 provides an example of a periodic task). In this example, it is also possible to note the use a real-time time base. The language provides in the specification both a wall calendar time base as well as a monotonic clock (in Annex D, dedicated to real-time mechanisms). Absolute time is supported through the `delay until` language keyword.

```ada
task body Cyclic is
   Next_Period : Ada.Real_Time.Time := First_Release;
   Period : Ada.Real_Time.Time_Span := Ada.Real_Time.Milliseconds(50);
   -- declarations
begin
   -- Initializations
   loop
      -- Task code
      Next_Period := Next_Period + Period;
      delay until Next_Period;
   end loop;
end Cyclic;
```

**Listing 7**. Ada periodic task.

```ada
protected type Bounded_Buffer is
    entry Get (X : out Item);
    entry Put (X : in Item);
private
   Get_Index : Buffer_Index := 1;
   Put_Index : Buffer_Index := 1;
   Count     : Buffer_Count := 0;
   Data      : Buffer_Array;
end Bounded_Buffer;

protected body Bounded_Buffer is
    entry Get (X : out Item) when Count > 0 is
    begin
        X := Data(Get_Index);
        Get_Index := (Get_Index mod Maximum_Buffer_Size) + 1;
        Count := Count - 1;
    end Get;
    entry Put (X : in Item) when Count < Maximum_Buffer_Size is
    begin
```

```ada
         Data(Put_Index) := X;
         Put_Index  := (Put_Index mod Maximum_Buffer_Size) + 1;
         Count := Count + 1;
      end Put;
end Bounded_Buffer;

My_Buffer: Bounded_Buffer;
```

**Listing 8**. Bounded buffer implemented with Ada protected objects.

Data sharing and synchronization is supported with protected objects, a language type which implements the monitor concept [33]. Listing 8 provides an example of the use of protected objects to define a type to support a bounded buffer (safe to use concurrently), while Listing 9 provides an example where a protected object is used to implement the mechanism for the release of a sporadic task.

```ada
protected Event is
  entry Wait;
  procedure Signal;
private
  Occurred: Boolean := False;
end Event;

protected body Event is
   entry Wait when Occurred is
   begin
      Occurred := False;
   end Wait;
   procedure Signal is
   begin
      Occurred := True;
   end Signal;
end Event;

task body Releasing is
begin
   -- ...
   if Some_Condition then
      Event.Signal;
   end if;
   -- ...
end Releasing;

task body Sporadic is
begin
   -- ...
   Event.Wait;
   -- Code
end Sporadic;
```

**Listing 9**. Task synchronization with Ada protected objects.

In the Real-Time annex (Annex D), the Ada language specification provides a very complete set of language features and libraries, intended to support real-time programming [32]. The scheduling model defined in this annex, allows for different scheduling policies to be used in the same application:

- `FIFO_Within_Priorities` – Within each priority level tasks are dealt with on a first-in-first-out basis. A task may pre-empt a task of a lower priority.

- `Non_Preemptive_FIFO_Within_Priorities` – Within each priority level to which it applies tasks run to completion until they are blocked or execute a delay statement. A task cannot be pre-empted by one of higher priority.

- `Round_Robin_Within_Priorities` – Whitin each priority level tasks are time-sliced with an interval that can be specified.

- `EDF_Across_Priorities` – This provides Earliest Deadline First dispatching. The general idea is that across a range of priorities levels, each task has a deadline and the one with the earliest deadline is processed first.

Since these policies are applied to priority levels (or range of priority levels), it is possible to define hierarchical scheduling (first level is priority-based) [34]. For instance, Listing 10 defines that the higher priorities (11 to 24) use fixed priority scheduling, tasks in the priority range 2 to 10 use dynamic scheduling, while background tasks are all placed in the lowest priority, using time-slicing.

```
pragma Priority_Specific_Dispatching(FIFO_Whitin_Priority,11,24);
pragma Priority_Specific_Dispatching(EDF_Across_Priorities,2,10);
pragma Priority_Specific_Dispatching(Round_Robin_Whitin_Priority,1,1);
```

**Listing 10**. Hierarchical scheduling.

Since both fixed and dynamic priorities are supported, the language also allows one to specify tasks properties such as priority and deadline (Listing 11).

```
task FPS_task
   with Priority => 20;

task EDF_task
   with Priority => 10,
        Relative_Deadline => Ada.Real_Time.Milliseconds(50);

protected Event
   with Priority => 20
```

**Listing 11**. Specification of priority and deadline properties.

As also noted in Listing 11, the language also supports assigning priorities to protected objects, in order to use priority inversion free data communication and synchronization between tasks, using the immediate ceiling protocol, together with the stack resource policy [35], or the Multiprocessor resource sharing Protocol (MrsP) [36]. All tasks are assigned a priority (for the case of tasks in the dynamic priority band, the priority defines the pre-emption level), and protected objects priorities can be set as a ceiling (the maximum of all tasks using the object). As soon as a task enters the protected object it will take the ceiling priority as its active priority. In the example, the protected object `Event` will be used to synchronize both `FPS_task` and `EDF_task`, therefore its priority needs to be at least the maximum of both tasks' priorities.

```ada
protected WCET_Overrun is
    entry Wait;
    procedure Fire;
private
    Occurred: Boolean := False;
end WCET_Overrun;

protected body WCET_Overrun is
    entry Wait when Occurred is
    begin
        Occurred := False;
    end Wait;
    procedure Fire is
    begin
        Occurred := True;
    end Signal;
end WCET_Overrun;

task body Periodic is
    Ex_Timer: Ada.Execution_Time.Timers.Timer(Current_Task);
begin
    loop
        Set_Handler(Ex_Timer,Task_WCET,
                    WCET_Overrun.Fire'Access);
        select
            WCET_Overrun.Wait;
            -- error recovery
        then abort
            -- task code
        end select;
        Cancel_Handler(Ex_Timer, Cancelled);
        Next := Next + Period;
        delay until Next;
    end loop;
end Periodic;
```

**Listing 12**. Use of execution time timers for WCET overrun.

The annex also provides support to the control of tasks' execution time (Listing 12), including the specification of execution time timers for tasks and interrupts, and task group budgets. Nevertheless, very few complete implementations of these mechanisms are freely available, except for those provided in the research oriented Marte OS operating system [37]. The approach in Listing 12 asynchronously calls `WCET_Overrun.Fire` if the timer expires, an alternative approach is to use the `Time_Remaining` function to check for overrun.

The annex provides the capabilities to partition multiprocessor implementations into disjoint dispatching domains, allowing for each domain to have its independent scheduling and dispatching queues. This allows one to pin tasks to cores or to setup limited migration approaches [38] by clustering tasks into domains (Listing 13).

It is also possible to assign tasks to CPUs dynamically during execution, e.g., depending on scheduling conditions, such as migrating a task after a certain execution time [39] (Listing 14).

```ada
-- CPUs start at 1
-- this code assumes there are 8 CPUs in the multiprocessor
Cluster_1: Dispatching_Domain := Create(1, 4);
Cluster_2: Dispatching_Domain := Create(5, 8);

task T1 with Dispatching_Domain => Cluster_1;

-- other tasks
```

**Listing 13**. Cluster-based scheduling.

```
if Release_Time >= Slot_Start and Release_Time < End_of_Phase_1 then
    Set_CPU(Client_Phase_1_CPU, Current_Task);
    Switch_Timer.Set_Handler(End_of_Phase_1, Handler'Access);
    Client_Current_Phase := Phase_1;
    Released := True;
elsif Release_Time >= Start_of_Phase_2
            and Release_Time < End_of_Slot then
    Set_CPU(Client_Phase_2_CPU, , Current_Task);
    Switch_Timer.Set_Handler(End_of_Slot, Handler'Access);
    Client_Current_Phase := Phase_2;
    Released := True;
else
    Client_Current_Phase := Not_Released;
    Switch_Timer.Set_Handler(Start_of_Phase_2, Handler'Access);
end if;
```

**Listing 14**. Implementation of a task splitting scheduling algorithm.

The specification of the Ada language includes numerous other mechanisms for concurrency and real-time. Nevertheless, in the late 90's there was a consensus that the Ada concurrency mechanisms were comprehensive but complex and could impair the use of the language model in critical systems. It was recognized that these systems also require simple and certifiable concurrency mechanisms. Namely, it was important to:

- Increase the efficiency of real-time kernels, by removing mechanisms with higher overheads.
- Simplify real-time kernels and remove non-certifiable mechanisms, to ease verification of high-integrity systems.
- Reduce non-determinism of applications by removing non-deterministic mechanisms.

To address this issue, the Ada real-time working group proposed to define a restricted profile of the concurrency mechanisms, leading to the Ravenscar [7] profile [40], which was standardized in the 2005 revision of the language. This profile defines a model based on fixed-priority scheduling, well accepted in the high-integrity/safety-critical industry, removing all dynamic concurrency mechanisms from the language. More recently, a slightly more relaxed profile is also defined, the Jorvik profile [41], but that maintains the same model of concurrency.

The use of a language with support for concurrency and real-time greatly improves the readability and correctness of applications, very important to cyber-physical systems and even more in critical applications. Nevertheless, some disadvantages also exist [24]. One is that many times different languages are used in different components of the application which is easier if an operating system specification is used. For instance, linking concurrent Ada and C code is much easier if the Ada runtime is implemented on top of a POSIX interface, with the C code using the same interface for concurrency. A second problem is the need to implement the language runtime on top of different platforms and operating systems. This means that there will be a time lag between the availability of a hardware or software platform, and the required language support [8].

---

[7] Although Ravenscar was in the beginning also taken as an acronym, the name comes from the small city in the north of England where took place the meeting of the International Real-Time Ada Workshop that defined the initial version of the profile.

[8] This is one of the issues with the use of the Ada language outside of the critical systems niche, as the cost of porting the language runtime to the multitude of existing processors and operating systems is high.

# 4. Programming Parallel Real-Time Systems

When moving from a concurrent model to both concurrency and parallelism, programming real-time systems needs to consider not only the issues listed above, but also the fact that the sequential code in the concurrent activities may be executed in parallel. It is necessary to combine in the programming model the capabilities to define the inherent concurrent activities of the application, and the potential parallelism in the activities' algorithms.

Moreover, and as noted in subsection 2.2, the thread scheduling needs to be augmented with the mapping of the parallel computation to the scheduling entities. Although this may seem to be an issue of the runtime and operating system, when the parallel and concurrent entities are specified at the programming model, mechanisms are required to also specify how the underlying runtime and operating system should manage mapping and scheduling (an integrated vertical stack as noted in Figure 4). Similar to the discussion in the previous subsections, this could be using library and calls to the underlying runtime, or directly supported at the programming model level.

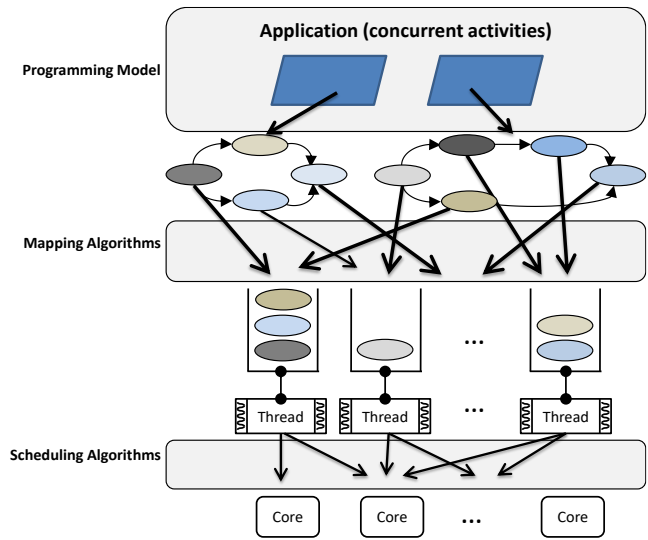

**Figure 4**. Vertical stack of mapping and scheduling parallel computation (adapted from [14]).

The programmability of parallel real-time systems is still a topic of research and development. Although a few approaches exist, there is no complete model (and, even less, stable products). The next sections of the paper provide an overview of two of the existent approaches, being currently researched and developed:

- The use of a subset of OpenMP [42], based on its tasking model, presented in Section 5.
- The Ada 2022 [43] parallel programming model integrated with the real-time features of the Ada language, in Section 6.

Although based on a similar fine-grained parallel model for real-time systems, these two approaches come from two different communities, and from opposing directions, which reflects in the main design differences and availability of mechanisms integrating parallelism and real-time. OpenMP comes from the high-performance community, supporting a very complete and complex parallel programming model. On the other hand, Ada is a technology mostly developed by the critical and real-time systems communities, providing extensive and complex support for programming real-time systems. Integrating real-time and parallelism implies (i) in OpenMP adding real-time to a parallel model, whilst (ii) in Ada adding a parallel model to a concurrent and real-time language.

Apart these two approaches, very little work exists in the integration of real-time and parallelism in other technologies. [44] provided an initial proposal to combine the Java fork//join model [45] with the Real-Time Specification for Java [28], while [46] did a preliminary analysis of real-time execution in the Embedded Multicore Building Blocks [47], Intel Threading Building Blocks [48] and High Performance ParalleX [49] frameworks. Nevertheless, to the best of the author's knowledge, these works have not been continued.

## 5. OpenMP Tasking Model

One of the frameworks which is presented to demonstrate its use for real-time applications is the use of the OpenMP tasking model [42] (Chapter 12), with precedence constraints defined through data dependencies. OpenMP is one of the most common parallel programming models in the high-performance computing domain, being increasingly used in embedded computing systems, as well as high-performance applications in critical computing systems (such as video processing in autonomous driving).

OpenMP originally considered a thread-centric model to extensively exploit data-parallel and loop-intensive types of applications. But since OpenMP 3.0, the specification has evolved into a task-centric model to allow for very complex types of irregular and fine-grained parallelism. Data dependencies were introduced in OpenMP 4.0, being intended to allow the implementation of a data-flow model of computation with task graphs. This model enables the programmer to define implicit and explicit tasks and data dependencies between them. Tasks are executed by a team of threads at runtime, which are managed by the underlying runtime, hiding complexity from the programmer.

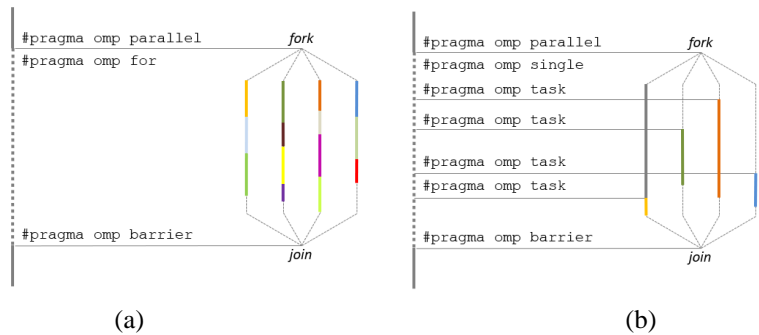

Figure 5. Fork/join model with (a) structured parallelism and (b) unstructured parallelism [14].

### 5.1 Support for Real-Time Parallel Models

An OpenMP program begins as a single thread of execution, called the initial thread. Parallelism is achieved through the parallel construct. When such a construct is found, a team of threads is spawned. These are joined when a barrier is encountered. Therefore, in its simplest form, the OpenMP tasking model fully supports the fork/join computational model (Figure 5). Within the parallel region, parallelism is achieved by means of different constructs: `for`, `section` or `task`. While `for` and `section` provide with structured parallelism, the tasking model full flexibility is achieved with unstructured parallelism.

The full flexibility of the OpenMP tasking model is achieved using both the nesting of task constructs, as well as data dependencies between tasks. In this case, the OpenMP tasking model can be used to program a parallel directed acyclic graph (DAG) [9] as presented in subsection 2, and shown in Listing 15 and Figure 6.

In order to understand the mapping between the OpenMP tasking programming model and the DAG real-time model, an important notion is the definition of OpenMP task scheduling points (TSP) [42] (section 12.9). TSPs are specific situations in the program where the task can be suspended, and the thread executing it can be allocated another task. TSPs occur upon task creation and completion, and at synchronization points such as `taskwait` directives, and explicit and implicit barriers. TSPs divide task regions into task parts executed uninterrupted (from the OpenMP point of view). Different parts of the same task region are executed in the order in which they are encountered, as it is a sequential execution.

The example shown in Listing 15/Figure 6 identifies the parts in which each task region is divided: `T0` is composed of parts `tp00` to `tp04`; `T1` is composed of parts `tp10` and `tp11`; `T2`, `T3` and `T4` are composed of a single part each. When the execution of a task reaches a TSP, the OpenMP runtime may continue with the current task, or resume a previously suspended task. How the runtime performs these decisions is left to the implementation, as long as it fulfils task scheduling constraints (TSC) [42] (section 12.9):

1. Scheduling of new tied tasks is constrained by the set of task regions that are currently tied to the thread and that are not suspended in a barrier region. If this set is empty, any new tied task may be scheduled. Otherwise, a new tied task may be scheduled only if it is a descendent task of every task in the set. This will be further discussed in subsection 5.3. [10].
2. A dependent task shall not start its execution until its task dependences are fulfilled.
3. A task shall not be scheduled while any task with which it is mutually exclusive has been scheduled but has not yet completed.
4. When an explicit task is generated by a construct that contains an if clause for which the expression evaluated to false, and the previous constraints are already met, the task is executed immediately after generation of the task.

---

[9] The most recent OpenMP specification also provides a Variant clause, which enables mapping to more recent DAG models in Real-Time Systems, intended for heterogenous platforms.
[10] In older versions of the OpenMP specification, this was TSC2, therefore many works can be found referring to it with this acronym.

```
#pragma omp parallel {
#pragma omp single                                                    // T0
{
        ...                     // tp00
        #pragma omp task depend(out: x)                               // T1
        {
            ...                 // tp10
            #pragma omp task if(false)                                // T4
            { ... }             // tp4
            ...                 // tp11
        }
        ...                     // tp01
        #pragma omp task depend(in: x)                                // T2
        { ... }                 // tp2
        ...                     // tp02
        #pragma omp taskwait
        ...                     // tp03
        #pragma omp task                                              // T3
        { ... }                 // tp3
        ...                     // tp04
}}
```

**Listing 15**. OpenMP structure example (adapted from [14]).

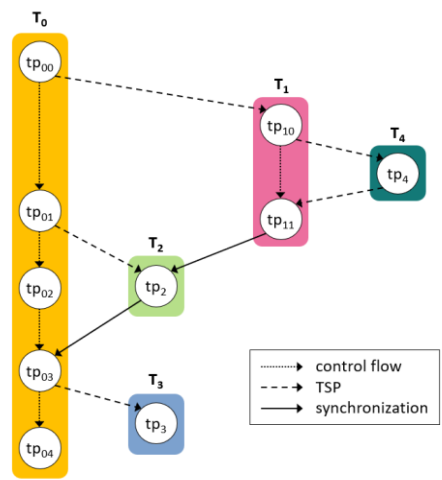

**Figure 6**. Example DAG (adapted from [14]).

The DAG in Figure 6 is then created by:

- Control flow dependencies between task parts of the same task (e.g., tp01 can only be executed after tp00).

- Task creation, since T1, T2 and T3 are created by T0, therefore, tp10 can only execute after tp00, tp2 only after tp01 and tp3 after tp03; T4 is created by T1, so tp4 is after tp10.

- Task completion, due to the TSC of the if clause, which forces T4 to be executed immediately when created, implying that tp11 waits for tp4.

- Synchronization due to TSC on data dependencies, as T2 depends (in) on variable x, which T1 writes (out), therefore tp2 depends on tp11.

- Synchronization due to barriers, the taskwait directive implies tp03 needs to wait for the completion of T2 (tp2), T1 (tp11) and T4 (tp4). Since tp2 depends on tp11 and tp11 depends on tp4, only the arrow between tp2 and tp03 is necessary.

A simplification may be used, with a restricted model using only data dependencies to specify the DAG edges. In this case, the program is developed with all tasks at the same level (except for the initial task of the single directive), and synchronization between tasks only using the depend clause of tasks (Listing 16 and Figure 7).

```c
void cholesky(float *M, int ntiles) {
   for (k = 0; k < ntiles; k++) {
      #pragma omp task depend(out:M[k][k])
      potrf (M[k][k]);
      for (i = k + 1; i < ntiles; i++)
         #pragma omp task depend(in:M[k][k]) depend(out:M[k][i])
         trsm (M[k][k], M[k][i]);
      for (i = k + 1; i < ntiles; i++) {
         for (j = k + 1; j < i; j++)
            #pragma omp task depend(in:M[k][i],M[k][j])
                             depend(out:M[j][i])
            gemm (M[k][i], M[k][j], M[j][i]);
         #pragma omp task depend(in:M[k][i])
                          depend(out:M[i][i])
         syrk (M[k][i], M[i][i]);
      }
   }
}
```

**Listing 16**. Structure with only data dependencies

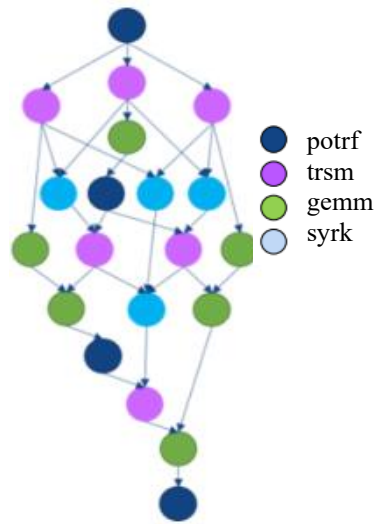

**Figure 7**. OpenMP DAG with data dependencies

Even with this simplification, the OpenMP DAG model provides for significant flexibility in the development of parallel real-time models. However, this flexibility comes with the additional burden to the programmer to build safe data sharing and synchronization between the tasks, an already difficult issue to consider in OpenMP [50].

Note that in order to be able to support offline real-time analysis, it is necessary that the full DAG is known offline. It is therefore necessary to extract all the information that impacts the parallel execution, i.e., tasks and data-dependencies whose instantiation depends upon loop iterations and/or if-conditions. If this is not possible to be done offline, then assumptions on pessimistic behaviour need to be considered, leading to pessimistic schedulability analysis [50][51].

Although enabling the use of real-time parallel models, the current specification of OpenMP does not support real-time scheduling properties, such as deadlines, time bases or worst-case execution time (WCET). Furthermore, and although supporting the notion of task priorities, its semantics are very relaxed (the clause specifies a "hint" [42] (section 12.4)), thus real-time behaviour cannot be guaranteed.

A proposal was made [52] to define an OpenMP profile for critical real-time systems, which would include:

- Tightening of the semantics of the priority clause of tasks, which would provide for priority driven task scheduling algorithms.
- Specification of a deadline property for tasks, allowing for dynamic scheduling algorithms.
- Specification of task event clauses, to enable the recurrent release of a task either by a time base, or by a sporadic event.

This proposal was not considered, being the decision to leave fur future work, being currently the real-time behaviour an implementation approach. This is briefly mentioned in [52], which also discusses the different alternatives on how OpenMP specifies independent thread teams for different parallel regions, something which we will discuss in subsection 5.3.

## 5.2  Communication and Synchronization

OpenMP provides a small set of mechanisms to allow safe sharing of data between parallel tasks. The `depend` clause is one of such mechanism as it allows the implementation to guarantee ordering with mutual exclusion between parallel writes/reads between tasks. However, if the goal is to simply guarantee mutual exclusion when accessing shared data, OpenMP provides mechanisms such as the `critical` construct [42] (section 15.2), or `locks` [42] (section 18.9). From a real-time perspective, and as OpenMP does not support the specification of real-time properties, these mechanisms do not implement any type of priority inheritance or ceiling [11].

An implementation of the OpenMP runtime may map these mechanisms to underlying POSIX mutexes, with the real-time extensions, as presented in subsection 3.1. However, there is no mechanism at the programming model that allows the programmer to control this behaviour. For this to be correctly used, it would be necessary to be programmed inside the OpenMP runtime.

## 5.3  Control of Mapping and Scheduling

The OpenMP tasking model provides a two-level mapping and scheduling approach, similar to the one previously provided in Figure 4. OpenMP tasks specified within the program are mapped to the team of threads of the parallel region, which are then executed by the operating system.

---

[11] Not related to real-time, but the use of critical constructs and locks in OpenMP tasks needs to be carefully designed, since these mechanism work at a thread level. This is particularly true with untied tasks, which can migrate between threads.

Concerning the mapping of tasks to threads, the specification also leaves it open to the implementation. Typical implementations use breadth-first (BFS) and work-first (WFS) approaches, where BFS creates all children tasks before executing them while WFS executes new tasks immediately after they are created (theoretically is better as it improves cache locality). Some implementations use a single queue for all tasks (in the same thread team), but more recent approaches (e.g., the LLVM implementation) use work-stealing strategies [19]:

- Each thread has its own queue, tasks created are inserted at the head of its thread queue.
- Threads first try to execute tasks from its queue (head first, to increase cache locality).
- If a thread's queue is empty, a thread will remove (steal) from another's thread queue, but this time form the tail, reducing contention and probability of impact on cache (older "cold" tasks are being taken).

OpenMP distinguishes between tied [12] (by default) and untied [13] tasks. A tied task is bound to be executed by the same thread, while an untied task may migrate from thread to thread, at task synchronization points. Untied tasks provide a work-conserving policy, where a thread can suspend execution of the current task and switch to executing another task. The suspended task can resume execution on the same thread or on a different thread depending on the task type, if available. It guarantees that no threads remain idle when there is a work to be performed, an important concept in real-time system scheduling, as it allows for simpler and less pessimistic analysis to be performed [53]. Also, for work-conserving, approaches exist for real-time friendly work-stealing implementations [18].

On the contrary, the use of tied tasks may lead to non-work-conserving task scheduling, due to the first task scheduling constraint presented in subsection 5.1. The reason for this constraint is a consequence of a trade-off between the thread-centric and task-centric specification of OpenMP. Because many OpenMP mechanisms work at a thread level (critical construct, thread private variables), allowing tasks to migrate among threads is a potential source of concurrency errors, therefore by default it is disallowed. Moreover, when tied tasks are created, they are forced to be mapped either in an empty thread, or a thread where all tasks are ascendant of the new task. Although analyses exist for this approach (e.g. [54]), these provide pessimistic results, in some cases with response times higher than the total execution time of the graph (thus worse than sequential execution). Note that the use of tied tasks with the WFS mapping approach will lead to sequential execution.

OpenMP also does not specify how the team threads are scheduled [14]. This makes it impossible to provide any real-time scheduling control at the programming level. Moreover, the specification makes teams from two parallel regions completely independent, and oblivious of the other teams. Although many real-time works consider the schedulability analysis of a single DAG, it is possible that an application is required to implement parallelism in more than one of its concurrent activities.

---

[12] All task-parts of a tied task must be executed by the same thread.
[13] Each task-part of an untied task can be executed by a different thread.
[14] Most implementations map the OpenMP threads 1-to-1 to operating system threads, such as pthreads.

This is noted in [52], which proposes alternatives using nested tasks or nested parallel regions, but that are dependent on the ability to specify real-time priorities or deadlines so that an OpenMP runtime manages the full scheduling of tasks and threads (Listing 17).

```
#pragma omp parallel
#pragma omp single
{
    #pragma omp task priority (p1) // T1 : OpenMP-DAG1
    { RT task 1 ( ) }
    #pragma omp task priority (p2) // T2 : OpenMP-DAG2
    { RT task 2 ( ) }
    // ...
    #pragma omp task priority (pn) // Tn : OpenMP-DAGn
    { RT task n ( ) }
}
```

**Listing 17**. Proposal of real-time OpenMP [52].

Another potential approach is to rely on the fact that implementations of OpenMP delegate to the underlying operating system the scheduling of the threads. This way, the real-time behaviour can be provided by an underlying real-time operating system. In this case, real-time concurrent activities could be mapped to different real-time threads (e.g., with pthreads), of different priorities, with each thread executing a different DAG (Listing 18). This approach has the additional advantage of allowing the development of applications using components in different languages, as long as all using the same underlying threading model.

```
void *T1(void *arg) {
    // T1 : OpenMP-DAG1
    #pragma omp parallel {
    #pragma omp single {
        // ...
    }}
    return NULL;
}

void *T2(void *arg) {
    // T2 : OpenMP-DAG2
    #pragma omp parallel {
    #pragma omp single {
        // ...
    }}
    return NULL;
}

// ...

void main(){
//...

pthread_attr_init(&custom_attr_fifo1);
pthread_attr_setschedpolicy(&custom_attr_fifo1, SCHED_FIFO);
fifo_param1.sched_priority = p1;
pthread_attr_setschedparam(&custom_attr_fifo1, &fifo_param1);
pthread_create(&thread_id1, &custom_attr_fifo1, T1, NULL);

pthread_attr_init(&custom_attr_fifo2);
pthread_attr_setschedpolicy(&custom_attr_fifo2, SCHED_FIFO);
fifo_param1.sched_priority = p2;
pthread_attr_setschedparam(&custom_attr_fifo2, &fifo_param2);
pthread_create(&thread_id1, &custom_attr_fifo2, T2, NULL);
// ...
```

**Listing 18**. Mixing pthreads with OpenMP.

However, for this approach to work, it is necessary that the underlying OpenMP runtime implementation propagates the priority of the thread creating the parallel region to the other threads of the team created in the parallel region. A similar approach was used in the implementation of the UpScale framework [55], which implemented a lightweight OpenMP runtime [56] that propagated thread priorities to the team threads and to the underlying operating system (in this case the Erika OS [57]). This implementation also propagated to the operating system the information on the task scheduling points. This allowed the use of a limited pre-emptive scheduling approach [58] in the operating system, reducing overhead and increasing cache locality [14].

Also necessary is the ability to control the affinity of threads to the underlying processors. OpenMP includes the `proc_bind` directive, that allows the programmer to specify the policy for assigning threads to cores. The provided mechanisms allow defining that all threads in the team are bound to the same core, or bound to consecutive cores, or spread across cores from different partitions (clusters). It is therefore possible to define global or partitioned scheduling, but not more flexible approaches such as cluster scheduling or task splitting [38][39]. Again, these need to be provided by a specific implementation.

The use of the OpenMP tasking model in real-time systems is still the subject of research. A limited implementation exists of some of the features in [52], but more work is still required, particularly on the integration of multiple parallel real-time activities in the same application, and how to control mapping of tasks to threads for more predictable (and less pessimistic) analysis [17].

## 6. Ada 2022

As presented in subsection 3.2, the Ada programming language provides extensive support to concurrency and real-time systems. Complementing this support, the forthcoming revision of the language [43] also includes a fine-grained parallelism model [59].

In this model, the `parallel` keyword is used to denote language constructs that can be executed in parallel by multiple threads of control [43] (section 5.1), both in `block` statements [43] (section 5.6.1) (Listing 19) as well as `loops` [43] (section 5.5) (Listing 20 and Listing 21). The specification also provides control of loop partitioning into chunks and reduction of variables (Listing 22).

The integration of the parallel model with the pre-existent concurrency model of Ada is provided by changing the notion of Ada tasks [43] (section 9) to allow an Ada task, when executing a `parallel` construct, to represent multiple logical threads of control which can proceed in parallel. Therefore, parallel blocks and loops can be used inside task bodies (Listing 23).

```ada
procedure Traverse (T : Expr_Ptr) is
begin
   if T /= null and then T.all in Binary_Operation'Class then
     parallel do
       Traverse (T.Left);
     and
        Traverse (T.Right);
     and
         Put_Line("Processing" &
                 Ada.Tags.Expanded_Name (T'Tag));
     end do;
   end if;
end Traverse;
```

**Listing 19**. Parallel block example [60] (section 5.6.1).

```ada
parallel for I in Grid'Range(1) loop
   Grid(I, 1) :=
        (for all J in Grid'Range(2) => Grid(I,J) = True);
end loop;
```

**Listing 20**. Parallel loop example [60] (section 5.5).

```ada
parallel for Element of Board loop
    Element := Element * 2.0;
end loop;
```

**Listing 21**. Parallel iterator example [60] (section 5.5.2).

```ada
declare
   subtype Chunk_Number is Natural range 1 .. 8;
   Partial_Sum,
   Partial_Max : array (Chunk_Number) of Natural :=
              (others => 0);
   Partial_Min : array (Chunk_Number) of Natural :=
              (others => Natural'Last);
begin
   parallel (Chunk in Chunk_Number)
   for I in Grid'Range(1) loop
      declare
         True_Count : constant Natural :=
            [for J in Grid'Range(2) =>
             (if Grid (I, J) then 1 else 0)]'Reduce("+",0);
      begin
         Partial_Sum (Chunk) := @ + True_Count;
         Partial_Min (Chunk) := Natural'Min(@, True_Count);
         Partial_Max (Chunk) := Natural'Max(@, True_Count);
      end;
   end loop;
   Put_Line
    ("Total=" & Partial_Sum'Reduce("+", 0)'Image & ", Min=" &
     Partial_Min'Reduce(Natural'Min, Natural'Last)'Image &
     ", Max=" & Partial_Max'Reduce(Natural'Max, 0)'Image);
end;
```

**Listing 22**. Parallel loop example with chunks and reduction [60] (section 5.5).

```ada
function Sum(T: Some_Array_T; First, Last: Some_Index_T)
    return Integer is
    S : Integer := 0;
begin
    parallel (Num_CPUs)
    for I in First .. Last when I mod 2 = 1 loop
        Atomic_Add (S, I);
    end loop;
    return S;
end Sum;

task body Cyclic is
   Next_Period : Ada.Real_Time.Time := First_Release;
   Period : Ada.Real_Time.Time_Span :=
           Ada.Real_Time.Milliseconds(500);
   T: Some_Array_T;
   S, S_L, S_R: Integer;
begin
   -- Initializations
   loop
      delay until Next_Period;
      -- ...
      parallel do
         S_L := Sum(T, T'First, T'First + T'Length/2 - 1);
      and
         S_R := Sum(T, T'First + T'Length/2, T'Last);
      end do;
      S := S_L + S_R;
      -- ...
      Next_Period := Next_Period + Period;
   end loop;
end Cyclic;
```

**Listing 23**. Task with fine-grained parallel execution.

## 6.1 Support to Real-Time Parallel Models

The parallel mechanisms provided by the forthcoming Ada language version, only provides support for the fork/join parallel model. Each parallel construct splits computation into a set of parallel threads of control, which are required to terminate so that the construct terminates. Figure 8 shows the instantiation of this model for the code in Listing 23, considering 4 CPUs.

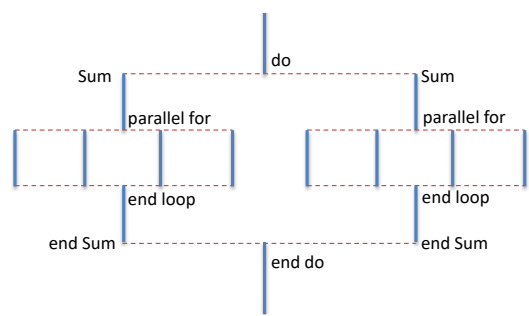

**Figure 8**. Fork/join model of Ada parallel computation.

Although a more general model was also considered [59], the underlying reasoning on Ada design is to allow the compiler to detect as much as possible errors at compile time, rather than waiting for these to emerge at run time, removing as much as possible the burden from the programmer to build safe concurrent (and parallel programs). It was decided that the new capabilities added to the language to support parallelism would also allow the compiler to detect as many such problems as possible, with reduced performance loss. This led to restricting the model to fork/join, as well as introducing new annotations to specify usage of global data and eventually blocking operations [61]. Performance is sacrificed, but only in very specific situations where allowing "un-joined" loop iterations (e.g., wavefront) across parallel loops provides faster processing [62].

The development of the parallel model of Ada maintained the possibility of limited compatibility with the OpenMP specification, in that the Ada model would be possible to be implemented on top of an OpenMP runtime [62][63]. This allows for applications being built with libraries in different languages such as C, C++ and Ada, all using the same OpenMP support for parallelism. It also allows an Ada program to make calls to the underlying runtime, using unstructured parallelism, but at the penalty of stepping outside of the safety box of Ada.

In the proposed approach for this integration, Ada tasks with no parallelism are mapped to one operating system thread, and each Ada task with parallel constructs is associated with a thread pool provided by OpenMP to execute the fine-grained parallelism (Figure 9). Additionally, it is assumed that each OpenMP thread is mapped 1-to-1 to operating system threads, which is typically the case. This simplifies analysis and implementation, as well as maintaining a consistent scheduling within the operating system.

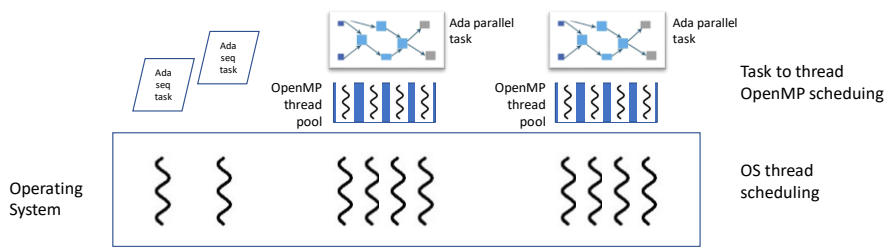

**Figure 9**. Mapping of Ada to OpenMP.

This approach is hindered because of independence and lack of semantics of multiple parallel teams in OpenMP (subsection 5.3), thus it is still an open area of work. An alternative approach provides templated execution that forces a unique OpenMP team of threads to be accessed from any Ada task [64], but that requires explicit changes to the program and may suffer from priority inversion (Figure 10).

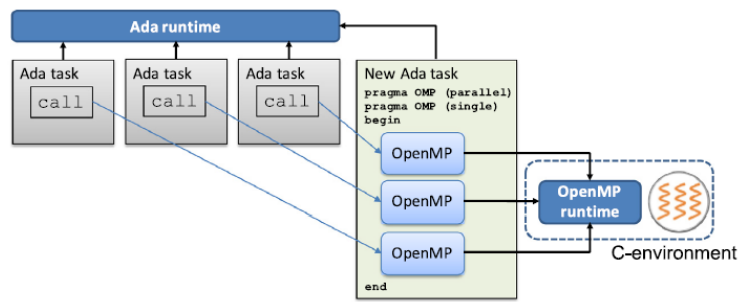

**Figure 10**. Integration of Ada and OpenMP with template execution [64].

As the parallel constructs in Ada integrate with the existing concurrency model, the Ada task properties for real-time scheduling (priority, deadline, execution time, etc), are still applicable, as well as the complete Ada scheduling model. Some clarifications and documentation are still necessary, e.g., when a high priority task pre-empts the parallel execution of a low-priority task, if the parallel computation is pre-empted in all cores, even if the high-priority task is not using them.

Moreover, as presented in Listing 23, it is also possible to define recurrent periodic real-time tasks with parallelism (and similar with aperiodic/sporadic tasks). Therefore, this model allows the programmer to easily build parallel real-time applications using the fork/join model, with an Ada runtime supporting both concurrency and parallelism, or with an OpenMP implementing the approach discussed in Listing 18.

## *6.2 Communication and Synchronization*

Since the parallel execution is integrated with the concurrency model, data sharing between parallel execution can be performed using protected objects (or `Atomics` as shown in Listing 23). Listing 24 provides an example, where a task spans parallel search in an array, which may update a counter in a protected object. The example also shows the use of the priority ceiling protocol to guarantee no priority inversion and deadlock free execution if other tasks use the object (`Ceiling_Priority` ≥ `Some_Priority`).

```ada
protected Counter
      with Priority => Ceiling_Priority;
is
      procedure Inc;
      function Get return Natural;
private
      Count: Natural := 0;
end Counter;
protected body Counter is
      procedure Inc is
      begin
            Count := @ + 1;
      end Inc;
      function Get return Natural is
      begin
            return Count;
      end Get;
end Counter;

task Execute with Priority => Some_Priority;
task body Execute is
-- Declarations
begin
-- Initializations
      -- ...

      parallel
for Element of Trace loop
            if Element = 0 then
                  Counter.Inc;
            end if;
end loop;
      -- ...
end Execute;
```

**Listing 24**. Data sharing in parallel constructs.

Note that the reverse situation, which is spawning parallel execution inside protected objects is not allowed. If a parallel construct is found when executing a protected subprogram each element of the parallel construct that would have become a separate logical thread of control executes on the logical thread of control that is performing the protected action, in an arbitrary order [43] (section 9.5.1). The reason for this restriction relates to the potential complexity of handling the computation in different threads whilst holding the lock on the object.

Synchronization between parallel logical threads of control is not possible, as potentially blocking operations are disallowed [43] (section 5.1). This is a consequence of the design of a safer model (may not be detected at compile time, unless blocking annotations are provided), as it is intended that all synchronization is provided by the fork/join execution.

## 6.3 Control of Mapping and Scheduling

The Ada standard specifies that each parallel segment (chunk) is provided with its own logical thread of control but does not specify how these logical threads are mapped to the underlying operating system, except that these have the same scheduling properties as the Ada task. In [65] a model was provided for the underlying execution behaviour, based on the notion of abstract executors, which carry the actual execution of Ada tasks in the platform. The goal of this abstraction was to provide the ability to specify the progress guarantees that an implementation (compiler and runtime) need to provide to the parallel execution, without constraining how such implementation should be done:

- Immediate progress – when cores are available, logical threads of control which are ready to execute can execute to completion in parallel (limited only by the number of free cores).
- Eventual progress – when cores are available, ready logical threads of control might need to wait for the availability of an executor, but it is guaranteed that one will become available so that the logical thread of control will eventually be executed.
- Limited progress – even if cores are available, ready logical threads of control might need to wait for the availability of an executor, and the runtime does not guarantee that one will be eventually available. This means a bounded number of executors, which may block when logical threads of control block [15].

This model is then shown to be applicable for real-time systems, where logical threads of control run-to-completion in the same executor where they have started execution, although the executor can be pre-empted by higher-priority (or nearer deadline) executors, or even the same priority/deadline if the task's budget/quantum is exhausted.

Together with the scheduling mechanisms defined in the Ada Real-Time Systems Annex, it allows the use of a global work-conserving scheduling approach [15], which can be analysed with current real-time analyses.

However, some of the mechanisms proposed were not considered for the parallel model and are still only applicable to Ada tasks with no fine-grained parallelism [66]. In particular, a few issues are still open:

- The semantics of execution time metrics and control, in the presence of parallelized execution.
- The semantics and effects of changing real-time attributes within parallel constructs are not defined, parallel changes may lead to undesired effects.

---

[15] Since the decision was to not allow blocking inside parallel computation, limited progress is currently not used.

- The model for the control of Ada task affinities and dispatching domains is applicable to tasks, thus there is no mechanism to specify how the parallel threads of control of one task are mapped to processors.
- Fine-grained control of chunking and mapping of chunks to the threads of control, to enable static specification of mapping.

The last two are particularly relevant, since at the programmer level there is currently no mechanism to control the mapping or the affinities of the parallel execution, not being possible to define a static mapping approach with partitioned scheduling. This is only possible by integrating the mapping and partitioning in the underlying runtime, similar to the approach in [14] for OpenMP. This is still an open topic of research.

Although some limited implementations exist for the parallelism features of Ada 2022 [63][67], work still needs to be done for the integration of these features with the existing real-time capabilities of the language, an essential factor to allow for the open challenges to be addressed.

## 7. Conclusions

This paper provided an overview and motivated the need for a solid background on real-time parallel programming paradigms. Real-time systems are a challenging domain of work, with a wide range of applications, bringing forward concurrency and time properties as foundational properties of systems. Programming real-time systems requires paradigms that are able to describe and manage these concepts, as first-class entities, and not as an afterthought. The development of real-time systems introduces a few specific challenges, among which the need that programming models and technologies support the specification of concurrent activities, timing-related properties, and allow the control of the application execution in the platform. This is now exacerbated with the support for fine-grained parallel execution.

Current research in this area comes from the intersection of activities of groups from different communities, such as high-performance computing, real-time and critical systems. The two different approaches analysed in the paper (OpenMP and Ada) are examples of this multidisciplinary work, with the different design decisions emanating from the community where the work started. OpenMP comes from the high-performance community, based on a complete and complex parallel programming model. Ada is from the critical and real-time systems communities, providing extensive and complex support for programming real-time systems.

The paper also identifies potential topics for future research, particularly the specification of real-time properties and the control of concurrent and parallel execution in a vertical stack, from the programming model to the underlying mapping and scheduling layers.

## Acknowledgements

This research has been co-funded by the European Union's FP7 and Horizon 2020 research and innovation programmes under grant agreements No 611016 (P-SOCRATES Project) and No 871669

(AMPERE project). The author would like to thank S. Tucker Taft and Tullio Vardanega for comments on an earlier version of the document.